# Updating the CSNS injector linac to 250 MeV with superconducting double-spoke cavities[*]

LI Zhi-Hui (李智慧)[1,2;1)]   FU Shi-Nian(傅世年)[2)]

1) Institute of Nuclear Science and Technology, Sichuan University, Chengdu, 610065, China
2) Institute of High Energy Physics, Chinese Academy of Sciences, Beijing, 100049, China

**Abstract** In order to update the beam power from 100 kW to 250 kW in China spallation neutron source (CSNS) Phase II, one of the important measures is to replace the 80 meters long beam transport line between the present 80 MeV linac injector and the RCS to another kind of acceleration structure. In this paper, we proposed a scheme based on 324 MHz double-spoke superconducting cavities. Unlike the superconducting elliptical cavity and normal conducting CCL structure, the double-spoke cavity belongs to TE mode structure and has smaller transvers dimension compared with that of TH mode one. It can work at base frequency as the DTL section, so that the cost and complexity of the RF system will be much decreased, and the behaviors of the beam dynamics are also improved significantly because of the low charge density and larger longitudinal acceptance. Furthermore, because of the relatively longer interactive length between charged particle and the electromagnetic field per cell, it needs relatively less cell numbers and it has larger velocity acceptance compared with the double frequency TH structures. The superconducting section consists of 14 cryomodules, each of which includes 3 superconducting cavities and a doublet. The general considerations on cavity and beam dynamics design are discussed and the main results are presented.

## 1 Introduction

Neutron scattering is a powerful means to probe the structure of the microscopic world and has become an indispensable method in the advanced researches in physics, chemistry, biology, life science, material science, new energy, as well as in applications [1]. China spallation neutron source (CSNS) consists of a H[-] linac and a proton rapid cycling synchrotron (RCS). It is designed to deliver a beam power of 100 kW in Phase I, and has the potential to upgrade to 500 kW in Phase II by raising the linac output energy and increasing the beam intensity [2]. It can provide users a neutron scattering platform with high flux, wide wavelength range and high efficiency.

For the Phase I of CSNS, the injector linac is composed of a H[-] ion source, low energy beam transport line, 3MeV radio frequency quadrupole accelerator, medium energy beam transport line and three Alvarez type drift tube Linac (DTL) cavities. It can accelerate H[-] particle with 15 mA beam current of to 80MeV. Between the injector linac and the rapid cycling synchrotron (RCS), there is a beam line with length of about 80 meters. In phase II, the beam line will be substituted with another accelerating structure section, with which the beam will be accelerated to 250 MeV, and the beam current will also be increased to 30 mA. The main parameters of the added section are shown in Table 1.

[*] Supported by the National Nature Sciences Foundation of China (11375122, 91126003) and China ADS project.
1)   E-mail: lizhihui@ihep.ac.cn; lizhihui@scu.edu.cn

Table 1 the main parameters of the added section of the CSNS Linac

| Parameter | Value |
|---|---|
| Energy range ( MeV ) | 80-250 |
| Peak current ( mA ) | 30.0 |
| Norm. RMS Transverse emittance ( $\pi$mm.mrad ) | 0.2 |
| Norm. RMS Longitudinal emittance ( $\pi$mm.mrad ) | 0.4 |
| Total length ( m ) | <80 |

The normal conducting coupled cavity linac (CCL) type cavity is the most widely used normal conducting acceleration structure for this energy range. Compared with the normal conducting cavity, the superconducting one has several excellent properties such as larger aperture, lower AC power consumer and so on. Furthermore it has achieved a great progress on the performance of the low and medium beta superconducting cavities and the relevant radio frequency superconductivity technologies, the superconducting cavities are widely adopted in the recent designs for similar linacs such as C-ADS, ESS, SNS, SPL, X-Project [3-7]. It can be foreseen that the low energy RF superconducting technology will be rapidly developed in China in the following years with the progress of the C-ADS project [3]. In this paper a design based on multi-gap superconducting spoke cavity to update the injector energy to 250 MeV will be introduced and the beam dynamics results will be presented.

## 2 Cavity design

There are two types of ion superconducting cavities that are suitable for the CSNS Phase II injector energy range, namely elliptical cavity and multi-spoke cavity. The elliptical cavity is more maturely developed and has successfully operated in the high energy part of SNS linac (160 MeV-1.0 GeV). But for the CSNS upgrading section, it is not the best choice because of the following reasons: 1) the input energy is only 80 MeV, which requires a type of elliptical cavity with geometry beta only about 0.45, and it is still a big challenge to build such an elliptical cavity and needs intensive R&D; 2) the 324 MHz working frequency of the DTL part is too low for elliptical cavity. The transverse dimension of 324 MHz elliptical cavity is almost 1 m, and it makes the mechanical stability of the cavity poor and very large cross section of the cryomodules. In order to keep the transverse dimension in a reasonable value (half meters), the frequency has to be doubled to 648 MHz at least, and this will inevitably increase the complexity and cost of the whole RF system, especially for a 250 MeV machine. This is also the main reason to abandon the normal conducting CCL structure; 3) the frequency doubling will also introduce difficulties in beam dynamics and may cause beam loss in the frequency transition region. On the other hand, because the spoke cavity is based on TE mode, its outer diameter is usually less than 0.5λ, not like TM mode cavity, with transverse dimension of about 0.9λ [8]. So the transverse dimension of a 324 MHz spoke cavity is almost the same as an elliptical cavity working at double frequency.

Furthermore, the interaction length of charged particles with electromagnetic field per cell for a 324 MHz spoke cavity is two times that of a 628 MHz elliptical, so in order to obtain a certain voltage, the spoke cavity will need fewer cells, and which will increase the cavity velocity acceptance and mode separation of the cavity. Fig. 1 shows the acceleration efficiency of different cavity types, which is defined as the ratio of effective voltage of particle with velocity β and the maximum effective voltage can obtain, and it is a function of cell numbers of the cavity. We can see, by double-spoke cavity, it needs only one type of cavity to cover the whole energy range and the acceleration efficiency within the energy range is greater than 80%. Though the multi-spoke cavity is still in the stage of research in labs, it has attracted intensive interesting in the past decades and the R&D results in labs are quite promising [9, 10, 11]. In china, the study on the spoke cavity has achieved a lot of progress in the past two years within the frame of CADS project. Based on these facts, a 324 MHz double-spoke cavity is proposed for the CSNS injector linac upgrading.

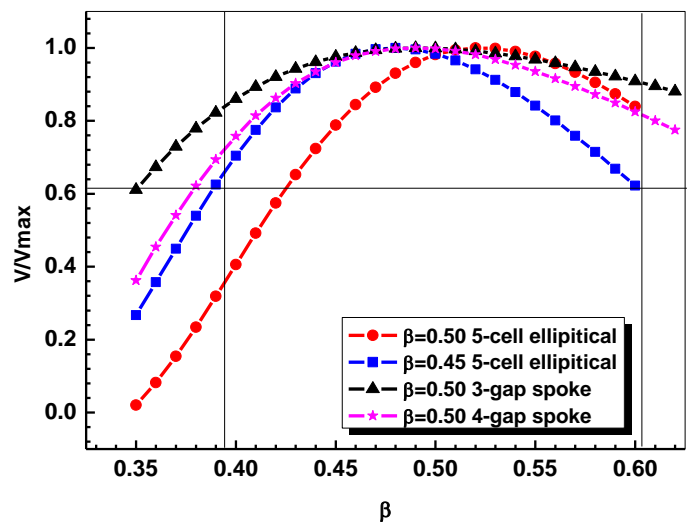

Fig. 1 The acceleration efficiency of different kind of cavity as function of particle velocity (The velocity range corresponding to the energy range is 0.38-0.62)

A double-spoke cavity with geometry beta of 0.5 is designed with the help of the CST Microwave package [12]. The geometry beta of the cavity is defined based on the distance between two spoke centers L, which is equal to the half of the production of RF wavelength and the geometry beta. The shape of the spoke is almost the same as the single spoke cavity for CADS, so that the fabrication of the double spoke cavity is nearly identical to two single spoke cavity with only one end panel, and the other side of the two cavities are annealed directly. The geometry of the cavity is shown in Fig. 2 and the main parameters are shown in Table 2.

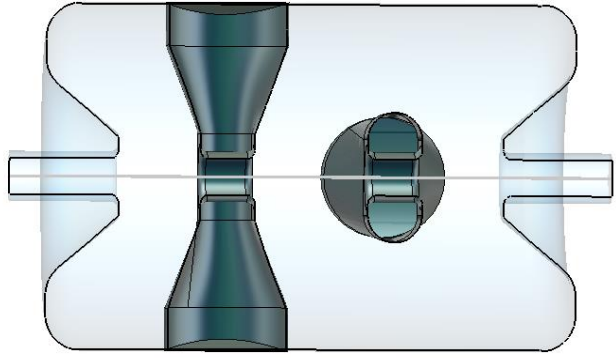

Fig.2 The geometry double spoke cavity with geometry beta 0.5

Table2 the main parameters of the double spoke cavity

| Frequency /$MHz$ | 324 |
|---|---|
| Aperture /$mm$ | 70 |
| Cavity length/$mm$ | 710 |
| Cavity diameter/$mm$ | 660 |
| Voltage ($E_{peak}$=30MV/m) /MV | 3.9 |
| $B_{peak}$ ($E_{peak}$=30MV/m) /mT | 73 |

From Table 2 we can see that the 3.9 MV maximum effective voltages can be obtained when the maximum surface electrical field is 30 MV/m and the corresponding surface magnetic field is 73 mT, and they are all in the reasonable range for technical feasibility [13]. If we take the average synchronous phase as 20 degree, then the energy gain per cavity is approximately 3.7 MeV, and the beam power will be about 110 kW for beam current of 30 mA, which also is a reasonable number for power couplers. Our final goal is to obtain 170 MeV energy gain within 80 m, so the average acceleration gradient should be greater than 2.125 MV/m. Given a filling factor of 0.5, the required acceleration gradient of the cavity should be greater than 4.25 MV/m, which is much smaller than the 5.2 MV/m, the maximum acceleration gradient of the cavity. So the parameters in Table 2 will be used as the base for the beam dynamic design.

## 3 Dynamics design

The main topics of the beam dynamics design of the superconducting section of the CSNS linac are the matching between the DTL and the superconducting section and the lattice design of the superconducting section, and both will be conducted following the widely accepted criteria for high current proton linac design [14]: (1) zero current phase advance per period less than 90 degree; (2) phase advance per meter change smoothly; (3) avoiding the possible dangerous resonances driven by space charge and parametric resonances.

The matching between DTL section and the superconducting section is performed with TraceWin code by varying the parameters of the triplet between them and of all the elements in the first two superconducting periods, so that the beam can be transformed smoothly to the matched shape as the acceptance of the superconducting section, at the same time the focusing strengths of both transverse and longitudinal directions are smoothly changed as the criteria (2) asked. The parameters of the DTL section will keep unchanged. Since the DTL design has already finished and in fabrication, so in order to satisfy the criteria (2), the phase advance per meter at the end of the DTL section is a very important parameter to determine the lattice structure of the superconducting section.

For the lattice design, the first important thing is to determine the ratio of the longitudinal focusing strength and that of the transverse one since it is a very important parameter to determine the stability of the dynamics. In our case it was determined by the DTL section design and is around 0.73. Now we have little freedom to change this ratio, but we can investigate if it is a proper number. We know one of the most important parametric resonances is the 1/2 resonance, which can be avoided in our case since we can keep the ratio of the focusing strengths of the transverse and longitudinal directions almost constant and far away from 0.5. The other kind of important resonances is driven by space charge, which is directly connected to the ratio between transverse and longitudinal focusing strengths and RMS emittances. The main possible resonance modes are usually plotted in the Hofmann chart as shown in Fig. 3. We can see, for tune depression greater than 0.6, which should be the situation in our case, the area marked with red ellipse is free from resonance and which is just the parameter region of the DTL section.

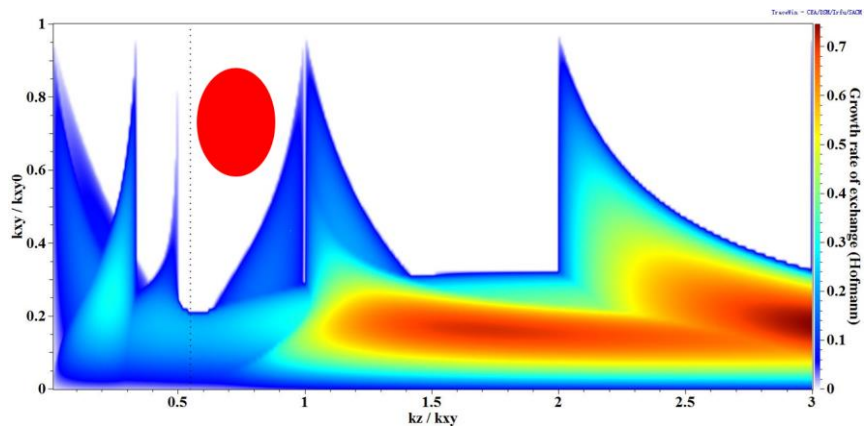

Fig.3 Hofmann Chart ($E_t/E_l$=2) drawn by traceWin

The lattice structure of the superconducting linac is determined by two aspects, one comes from the requirements of the techniques, which means sufficient longitudinal space should be kept for the installation of the RF cavities, the diagnostic devices, and focusing elements and the requirements of cryogenic system. The main parameters for the elements distribution are described as following. The distance between two cavities is reserved as 300 mm for the installation of helium tank and the tuner after reference the design of ProjectX [15] and ADTF [16], in which 240 mm and 300 mm are reserved, respectively; 500 mm on both sides of the cryomodules are reserved for temperature transition. Because the quadrupoles are used as transverse focusing elements in the upstream DTL sections, it is a natural choice to use quadrupoles as transverse elements in the superconducting section. The room temperature quadrupole doublet is placed between adjacent two cryomodules, and the space between two quadrupoles is 300 mm, which is reserved to house the diagnostic devices. The last problem required to be solved is to determine the number of cavities per period and this will be determined with the focusing strengths ratio and the criteria 1) and 2).

As criteria 2) shows, the maximum phase advance per period is 90 degree. After considering the longitudinal and transverse focusing strengths ratio 0.8, we can get the maximum longitudinal phase advance per period will be limited as 72 degree. Furthermore, the phase advance per meter should be as close as 18 degree, the value at the end of the DTL section. From these numbers we can find the period length should be about 4.0 m. With the parameters defined above, we get the final lattice structure as shown in Fig. 4. There are three cavities per period and the total length of the period is 4.33 m, and the length of the cryomodule is 3.73m. The zero current phase advances per meter and the synchronous phases are shown in Fig. 5. We can see the longitudinal phase advance is smoothly changed from DTL to superconducting section, but for the transverse one, there is a bulk located at the matching point which mainly comes from the focusing structure difference between DTL and the superconducting section and we can look at it as a section of beam line.

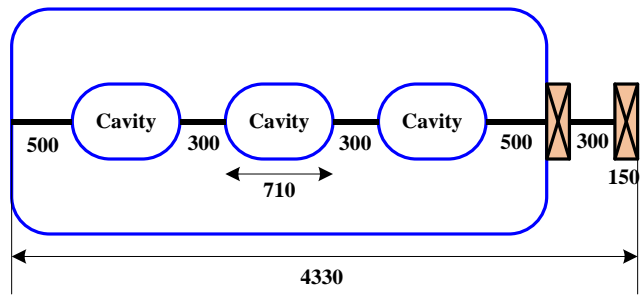

Fig. 4 Layout of the lattice structure

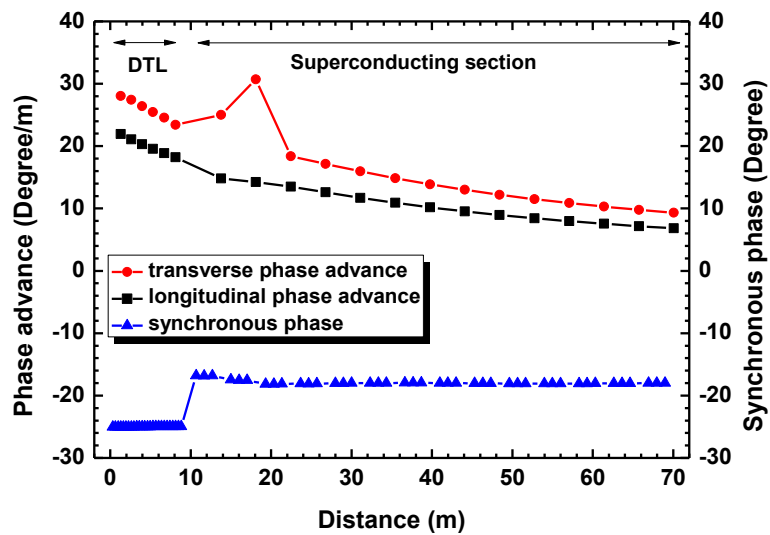

Fig.5 Phase advance per meter and synchronous phase

The parameters of the superconducting section are summarized in table 3.

Table 3 the main parameters of the superconducting section

| | |
|---|---|
| Total length (m) | 61.72 |
| Cavity number | 42 |
| Cryomodule number | 14 |
| Quad. number | 28+3(matching) |

| Lattice structure | R$^3$FD |
|---|---|

## 4 Multi-particles simulations

The multi-particle simulations are performed with TraceWin [17], which is one of the most famous high current linac design and simulation code developed by the CEA Saclay with functions such as automatic matching and matching parameters searching, error analysis, three dimension electromagnetic field map elements, 2D and 3D particle in cell space charge force calculation routings and so on. The simulation is started at the beginning of the last DTL cavity to the end of the superconducting section, and the parameters of the DTL cavity and the beam twiss parameters and emittances are provided by the designer of the 80 MeV CSNS linac injector, with which a 4σ truncated Gaussian initial particle distribution is produced by TraceWin with 100 thousand macro particles. The DTL cavity is modeled by the DTL cell elements provided by the TraceWin, and the superconducting cavities are modeled by the 3D field map based on the CST calculated three dimensional electromagnetic field distribution of the spoke cavity, while all the quadrupole magnetics are modeled with transmission matrix based on hard edge approximation and the validation is verified in the previous work [18].

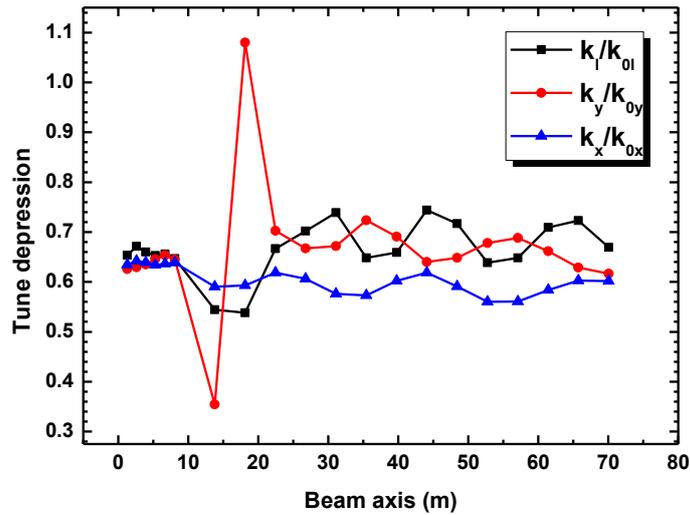

Fig. 6 Tune depressions along the last DTL cavity and the superconducting section

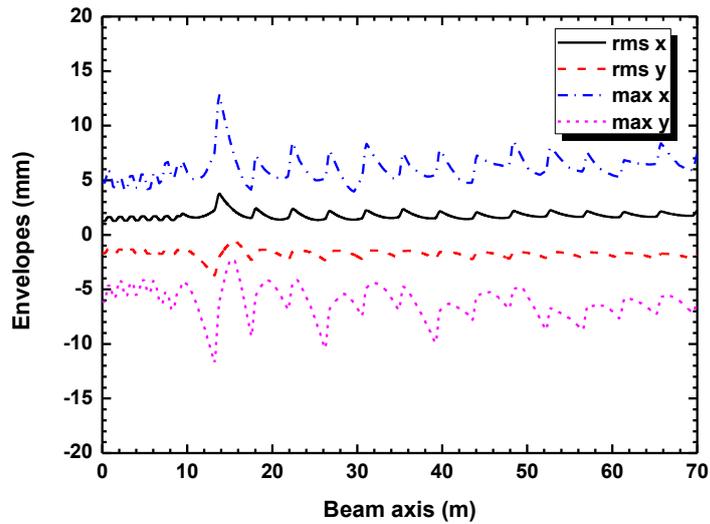

Fig.7 Envelopes evolution along the superconducting section

Fig. 6 shows the tune depressions along the linac, we can see they vary around 0.6 just as we predicted in the previous section, and it shows that the space charge effect should be given enough attention to this linac design. The rms and total envelopes of the beam along the linac are shown in Fig. 7. We can see the rms beam size is about 2 mm in x and y directions and the maximum beam size is less than 10 mm in the most part of the linac, and only in the matching section it is arrived almost 15 mm. In our case, the aperture of the cavity and the quadrupole is 30 mm and 35 mm, respectively, which are much larger than the maximum beam size. The total and rms phase width along the linac are shown in Fig. 8. The maximum phase is $\pm 10$ degree, much smaller than the bucket size with synchronous phase of -20 degree. Fig. 9 shows the emittance growth in all three phase planes along the linac, which is less than 20% and caused mainly by the charge redistribution at the beginning part of the simulation and the imperfect matching between DTL and superconducting section, which is hard to obtain when space charge effect presents. Finally, the phase space plots at the end of the linac are shown in Fig. 10 and the free of resonance design is proven by the relatively clear boundaries and regular shapes.

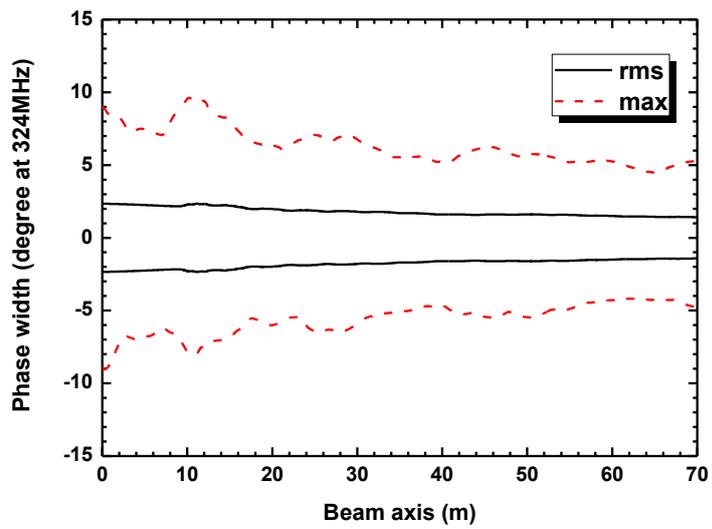

Fig.8 Phase width variation along the linac

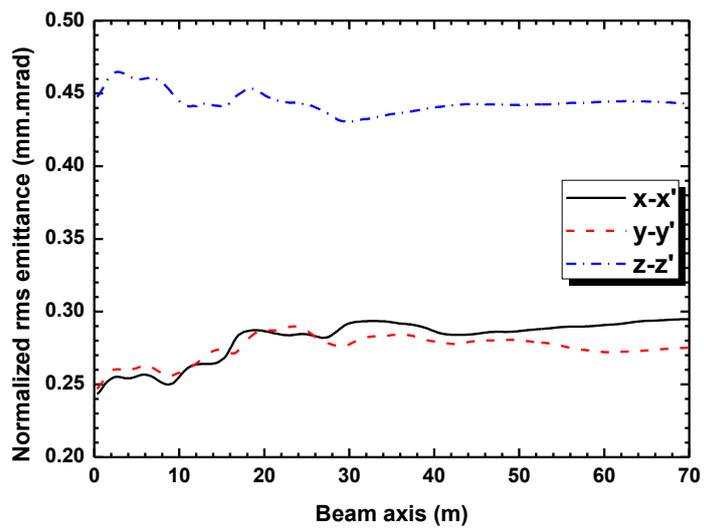

Fig.9 Normalized rms emittances growth along the linac

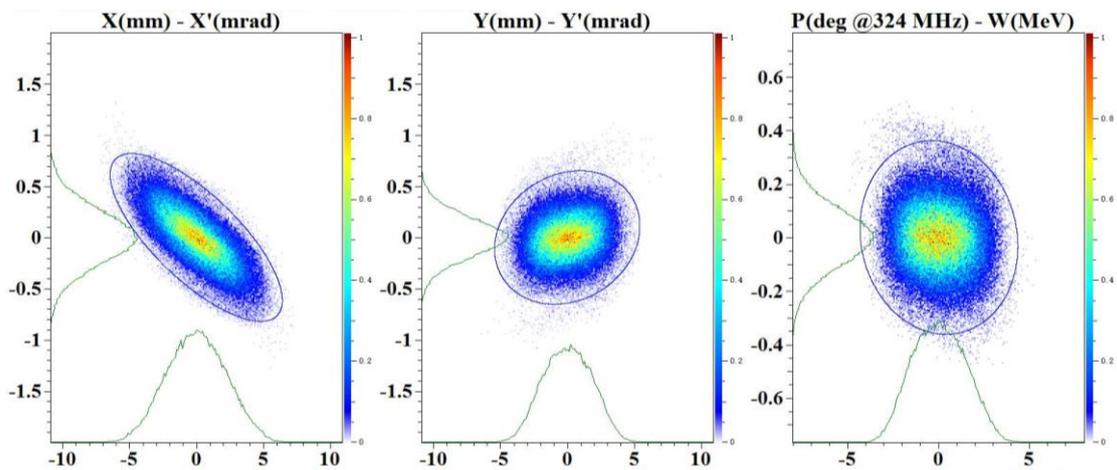

Fig. 10 Phase space distribution at the exit of the linac

We also investigated the beam dynamics stability under the errors, such as initial beam mismatches, the element alignment errors and so on. In order to study the effects of mismatches on the beam dynamics, we excited the quad mode, the fast mode and slow mode with 50% mismatch on each plan. The results show that with such large mismatches, the maximum beam sizes are almost doubled, but because of the relatively large aperture, no particle loss happens. The alignment errors listed in Table 4 are also studied, the results show that with a pair of steering and BPM per period, the residual center orbit can be controlled within 1 mm and with total number of $10^8$ particles, there is still no particle loss. The details of the error analysis will be presented in another paper.

Table 4 error settings

| | |
|---|---|
| Quadrupole transition errors (x/y/z mm) | 0.1/0.1/0.5 |
| Quadrupole rotation errors (x/y/z mrad) | 2/2/2 |
| Quadrupole strength errors (%) | 0.5 |
| Cavity transition errors (x/y/z mm) | 1.0/1.0/0.5 |
| Cavity rotation errors (x/y/z mrad) | 2/2/2 |
| RF field strength errors (%) | 1 |
| RF phase errors (degree) | 1 |

## 5 Conclusion

With superconducting double-spoke cavity, the output energy of CSNS injector linac can be extended to 250 MeV without frequency doubling. It needs 14 cryomodules and with three cavities in each. The total length of the superconducting section including the matching section is about 60 meters. The multi-particle simulations show that with 30 mA peak current, even with errors of beams and elements, there is no particle loss and proves that the proposed scheme can totally fulfill the requirements of the CSNS Phase II injector linac upgrading.

## Acknowledgement

The authors thank Dr. Peng Jun for her kindly providing the DTL design data.

## References

[1] SNS Brochure (09-G0046), http://neutrons.ornl.gov/media/pubs/
[2] J. Wei et al., Nuclear Instruments and Methods in Physics Research A 600 (2009) 10–13
[3] Zhihui Li et al., Physical Review Special Topics-Accelerators and Beams, 16, 080101(2013).
[4] M. Lindroos et al., Proceedings of LINAC2012, Tel-Aviv, Isreal, TH2A01
[5] J.Stovall et al., Proceedings of LINAC2000, Monterey, California, TUD22
[6] M. Vretenar, Conceptual deisgn of the SPL, a high-power superconducting H- lianc at CERN, CERN 2000-012
[7] N. Solyak et al., Proceedings of IPAC2010, Kyoto, Japan, MOPEC082
[8] C.S.Hooper and J.R.Delayen, Physical review special topics-accelerators and beams, 16,

102001 (2013)

[9] J.R.Delayen, Proceedings of LINAC2010, Tsukuba, Japan, TU302

[10] E.Azplaitin et al., Proceedings of EPAC2006, Edinburgh, Scotland, MOPCH158

[11] K.W.Shepard, M.P.Kelly, J.Fuerst, M.Kedzie and Z.A.Conay, Physica C, 441 (2006) 205-208

[12] https://www.cst.com/

[13] Hasan Padamsee, RF superconductivity, science, technology, and applications, Wiley-VCH verlag GmbH & Co. KGaA, Weiheim. 2009, p41-83

[14] F. Gerigk, Space-charge and beam halos in proton linacs, in: Proceedings of the Joint USPAS-CAS-Japan-Russia Accelerator School, Long Beach, California, 2002.

[15] I.Gonin et al., Proceedings of IPAC2010, Kyoto, Japan, WEPEC057

[16] J.P.Kelley et al., ADTF spoke cavity cryomodule concept, LA-UR-01-3818

[17] http://irfu.cea.fr/Sacm/logiciels/index3.php

[18] J.Y. Tang and Z.H. Li, edited, "Conceptual Physics design on the C-ADS accelerator", IHEP-CADS-Report/2012-01E.